\documentclass[superscriptaddress,aps,prb,reprint,twocolumn,amsmath,amssymb]{revtex4-2}
\usepackage{graphicx}
\usepackage{hyperref}
\usepackage{amssymb}
\usepackage{slashed}
\usepackage{dcolumn}
\usepackage{amsmath}
\usepackage{bm}
\usepackage{colordvi}
\usepackage{algorithm}
\usepackage{algpseudocode}
\usepackage{multirow}
\usepackage{titlesec}
\usepackage{newtxtext,newtxmath}
\usepackage{dsfont}
\usepackage{xcolor}
\usepackage{mathbbol}

\usepackage{hyperref}
\hypersetup{
    colorlinks=true,
    linkcolor=blue,
    citecolor=blue,     
    urlcolor=blue,
}

\allowdisplaybreaks

\usepackage{mathrsfs}
\makeatletter

\newcommand{\Rmnum}[1]{\expandafter\@slowromancap\romannumeral #1@}
\makeatother

\begin{document}

\title{Terahertz Control of Optical Second-Harmonic Generation in Displacive Ferroelectrics: Electronic Bloch-State Reconstruction}

\author{Hong-Kui Liu}
\affiliation{School of Information Science and Technology, University of Science and Technology of China, Hefei, Anhui 230026, China}

\affiliation{Guilin Julian Technology Co., Ltd., Guilin, Guangxi 541004, China}

\affiliation{School of Electronic Engineering and Automation, Guilin University of Electronic Technology, Guilin, Guangxi 541004, China}

\author{Zi-Chen Qin}
\email{202509040219@sust.edu.cn}
\affiliation{School of Chemistry and Chemical Engineering, Shaanxi University of Science and Technology, Xi'an, Shaanxi 710021, China}

\author{Jun-Song Wu}

\affiliation{Guilin Julian Technology Co., Ltd., Guilin, Guangxi 541004, China}

\affiliation{School of Electronic Engineering and Automation, Guilin University of Electronic Technology, Guilin, Guangxi 541004, China}

\author{Yue Yuan}
\email{yueyuanyy@mail.ustc.edu.cn}
\affiliation{Laboratory of Quantum Information, University of Science and Technology of China, Hefei, Anhui 230026, China}

\begin{abstract}
Optical second-harmonic generation (SHG) is a powerful probe of ferroelectric order, yet its microscopic origin and dynamical control are often understood primarily from symmetry considerations rather than from the underlying electronic processes. Here, we develop a microscopic theory of terahertz-controlled optical SHG in displacive ferroelectrics, establishing a direct connection between THz-driven polar lattice distortions and the resulting electronic nonlinear optical response. Starting from a complete Bloch-band representation, we show that an inversion-breaking lattice distortion reconstructs electronic Bloch wave functions, modifies optical dipole matrix elements, and activates nonlinear optical pathways that are forbidden in the centrosymmetric structure. We derive the second-order susceptibility in terms of the distortion-induced reconstruction of electronic states and optical transition matrix elements, and demonstrate that the electronic SHG susceptibility is linear in the polar distortion, $\chi^{(2)}({\bf Q})\propto{\bf Q}$, leading to an SHG intensity quadratic in the inversion-breaking order parameter. When the polar mode is coherently driven by a terahertz electric field, the resulting time-dependent lattice distortion dynamically reconstructs the electronic states and thereby modulates the optical SHG response, with $I_{2\omega}(t)\propto|{\bf Q}(t)|^2$ to leading order. This framework distinguishes the THz-driven lattice dynamics from the electronic interband processes responsible for optical SHG, which is particularly important in insulating ferroelectrics where low-energy carrier dynamics are absent. Our theory thus provides a microscopic bridge between nonequilibrium polar lattice dynamics and ultrafast electronic nonlinear optics.
\end{abstract}

\maketitle

\section{Introduction}
Ferroelectric materials exhibit strong second-harmonic generation (SHG) owing to their broken inversion symmetry, making SHG one of the most widely used probes of ferroelectric order~\cite{vonHoegen2018,Grishunin2017,Miyamoto2018,Zu2024,Zu2022,Abdelwahab2022,Dhongade2024}. Owing to this unique sensitivity, SHG has been extensively employed to identify ferroelectric phase transitions, image ferroelectric domains, determine polarization orientations, and investigate ultrafast polarization dynamics~\cite{li2023terahertz,cheng2023terahertz,74d5-4hsw}. In nearly all of these applications, however, the existence of SHG is interpreted primarily from symmetry considerations~\cite{vonHoegen2018,Grishunin2017,Miyamoto2018,Zu2024,Zu2022,PhysRevB.110.054311,7wnn-j6nc}: once inversion symmetry is broken, a finite second-order susceptibility is allowed by group theory.

Symmetry analysis establishes whether SHG is allowed, but does not reveal the microscopic mechanism by which ferroelectric distortions generate nonlinear optical responses. Ferroelectricity is governed by lattice distortions with characteristic energy scales of only a few meV~\cite{rowley2014ferroelectric,Yang2026,cowley1980structural,cochran1981soft,cochran1961crystal,cowley1996phase,cowley1965theory,cochran1969dynamical,cochran1960crystal,Yang2025,Yang2026PDW}, whereas optical SHG experiments typically involve photon energies of 1–3~eV~\cite{li2023terahertz,cheng2023terahertz,74d5-4hsw}, nearly three orders of magnitude higher. Moreover, the optical field oscillates on a timescale much faster than ionic motion, preventing the lattice from directly participating in the optical transition. Therefore, the observed SHG response is primarily electronic in origin~\cite{Haug1994,kittel1963quantum,PhysRevB.52.14636}. The ferroelectric lattice distortion is therefore expected to alter optical dipole matrix elements and virtual excitation pathways, which ultimately determine the nonlinear optical susceptibility.

Understanding how inversion-breaking lattice distortions control electronic SHG is therefore essential not only for clarifying the microscopic origin of ferroelectric nonlinear optics, but also for developing strategies to dynamically manipulate optical responses. In particular, ultrafast approaches such as THz-pump optical-probe spectroscopy provide a powerful route to coherently drive polar lattice distortions and thereby control electronic states and nonlinear optical functionalities~\cite{li2023terahertz,cheng2023terahertz,74d5-4hsw,Basini2024,Nova2019,Fechner2024,RevModPhys.93.041002,PhysRevB.74.155106}. However, establishing a microscopic relationship between a time-dependent lattice distortion and the resulting nonlinear optical response remains challenging. First-principles calculations can quantitatively evaluate nonlinear susceptibilities for specific crystal structures, but they generally do not provide an analytical framework connecting lattice dynamics, Bloch-wave-function reconstruction, optical transition matrix elements, and nonlinear optical responses.

Existing phenomenological descriptions~\cite{Ross2026,Ross2026KNbO3,PhysRevB.111.085109,Li2026THzNanotwin} introduce an effective electronic polarization as an additional dynamical variable coupled to the lattice. While such approaches can provide useful symmetry-level descriptions, the microscopic electronic polarization at optical frequencies originates from interband transitions and is therefore governed by fermionic Bloch states and optical matrix elements. A microscopic theory should therefore connect the THz-driven lattice coordinate directly to the reconstruction of electronic Bloch states and, consequently, to the nonlinear optical susceptibility, rather than treating the electronic polarization itself as an independently driven dynamical degree of freedom. This distinction is particularly important in insulating ferroelectrics, where low-energy electronic carrier dynamics are absent and the optical response originates predominantly from virtual interband transitions.

In this work, we develop a microscopic theory of ferroelectric SHG that explicitly establishes this connection, offering an effective theory for THz-pump optical-SHG-probe in displacive ferroelectrics. Starting from a microscopic electronic Hamiltonian in a complete Bloch-band basis, we derive an analytical expression for the second-order susceptibility induced by inversion-breaking lattice distortions. We show that polar distortions reconstruct Bloch wave functions, modify optical dipole matrix elements, and activate previously forbidden nonlinear optical pathways, resulting in a susceptibility that is linear in the inversion-breaking distortion. This framework provides a microscopic understanding for ferroelectric SHG and offers a basis for predicting and controlling nonlinear optical responses under equilibrium and nonequilibrium lattice distortions, including THz-driven ultrafast manipulation.

\section{All-band formulation of electronic SHG}
\label{dwdwd}

To establish a microscopic theory of electronic SHG that is directly applicable to realistic materials, we formulate the
problem in a complete Bloch-band basis. The present formulation explicitly includes all occupied,
unoccupied, and virtual intermediate electronic states. The virtual optical
process is obtained by systematically eliminating the intermediate states
through a block-diagonalization procedure.

\subsection{Hamiltonian}

We first consider the electronic Hamiltonian in the centrosymmetric
reference structure~\cite{kittel1963quantum,Haug1994},
\begin{equation}
H_0=
\sum_{m\mathbf{k}}
\varepsilon_{m\mathbf{k}}
|u_{m\mathbf{k}}^{(0)}\rangle
\langle u_{m\mathbf{k}}^{(0)}|,
\end{equation}
where $m$ denotes a complete set of Bloch bands and $|u_{m\mathbf{k}}^{(0)}\rangle$ are the eigenstates in the absence of
inversion-symmetry breaking distortion. The inversion-breaking structural distortion is described by a polar order
parameter ${\bf Q}$~\cite{rowley2014ferroelectric,Yang2026,cowley1980structural,cochran1981soft,cochran1961crystal,cowley1996phase,cowley1965theory,cochran1969dynamical,cochran1960crystal,Yang2025,Yang2026PDW,3y1m-66s1},
\begin{equation}
    {\bf Q}=(Q_x,Q_y,Q_z),
\end{equation}
which represents the polar lattice displacement associated with the
inversion-breaking phonon mode. Under spatial inversion,
\begin{equation}
    {\cal P}:{\bf Q}\rightarrow-{\bf Q}.
\end{equation}
The electronic states therefore become  
$|u_{m\mathbf{k}}\rangle
=
|u_{m\mathbf{k}}({\bf Q})\rangle $ as functions
of ${\bf Q}$. The coupling between electrons and an optical electric field is~\cite{Haug1994} 
\begin{equation}
H_{\rm int}(t)
=
-\hat{\mathbf d}\cdot {\bf E}(t).
\end{equation}
The corresponding matrix elements are 
$V_{mn}(\mathbf{k},t)
=
-d_{mn}^{i}(\mathbf{k},{\bf Q})
E_i(t)$, where $d_{mn}^{i}(\mathbf{k},{\bf Q})
=
\langle
u_{m\mathbf{k}}({\bf Q})
|
\hat d_i
|
u_{n\mathbf{k}}({\bf Q})
\rangle$,

We now analyze how the vector polar distortion ${\bf Q}$ enters the optical
matrix elements. To leading order, the lattice distortion generally leads to a lattice potential~\cite{kittel1963quantum}, 
\begin{equation}
    H_{\bf Q}
    =
    {\bf Q}\cdot{\bf U}
    =
    \sum_{\alpha=x,y,z}Q_\alpha U_\alpha ,
\end{equation}
where ${\bf U}=(U_x,U_y,U_z)$ is the vector of electronic perturbations
generated by polar displacements along the corresponding Cartesian directions. Then, the electronic eigenstates in the distorted structure can be
expanded around the centrosymmetric reference states~\cite{sakurai2020modern}:
\begin{equation}
    |n,\mathbf k;{\bf Q}\rangle
    =
    |n,\mathbf k;0\rangle
    +
    {\bf Q}\cdot
    |\boldsymbol{\delta} n,\mathbf k\rangle
    +
    O(Q^2),
\end{equation}
where 
    $|\boldsymbol{\delta} n,\mathbf k\rangle
    =
    \sum_{m\neq n}
    \frac{
    \langle m,\mathbf k;0|
    {\bf U}
    |n,\mathbf k;0\rangle
    }
    {
    \varepsilon_n(\mathbf k)-\varepsilon_m(\mathbf k)
    }
    |m,\mathbf k;0\rangle$. Here
$\langle m,\mathbf k;0|{\bf U}|n,\mathbf k;0\rangle$
is a vector in polar-distortion space. The dipole matrix element becomes
\begin{align}
    d_{mn}(\mathbf k,{\bf Q})
    &=
    \langle m,\mathbf k;{\bf Q}|
    \hat{\mathbf d}\cdot\hat{\mathbf e}
    |n,\mathbf k;{\bf Q}\rangle\nonumber\\
   &=
    d_{mn}^{(0)}(\mathbf k)
    +
    {\bf Q}\cdot
    {\bf d}_{mn}^{(1)}(\mathbf k)
    +
    O(Q^2),
\end{align}
where 
    $d_{mn}^{(0)}(\mathbf k)
    =
    \langle m,\mathbf k;0|
    \hat{\mathbf d}\cdot\hat{\mathbf e}
    |n,\mathbf k;0\rangle$, 
and
\begin{align}
    {\bf d}_{mn}^{(1)}(\mathbf k)
    &=
    \langle \boldsymbol{\delta} m,\mathbf k|
    \hat{\mathbf d}\cdot\hat{\mathbf e}
    |n,\mathbf k;0\rangle
    +
    \langle m,\mathbf k;0|
    \hat{\mathbf d}\cdot\hat{\mathbf e}
    |\boldsymbol{\delta} n,\mathbf k\rangle\nonumber\\
    &=
    \sum_{\ell\neq m}
    \frac{{\bf U}_{m,l,{\bf k}}d_{\ell n}^{(0)}(\mathbf k)
    }
    {
    \varepsilon_m(\mathbf k)-\varepsilon_\ell(\mathbf k)
    }
    +
    \sum_{\ell\neq n}
    \frac{ d_{m\ell}^{(0)}(\mathbf k){\bf U}_{l,n,{\bf k}}
    }
    {
    \varepsilon_n(\mathbf k)-\varepsilon_\ell(\mathbf k)
    }.
\end{align}
Here, ${\bf U}_{m,l,{\bf k}}=
    \langle m,\mathbf k;0|{\bf U}|\ell,\mathbf k;0\rangle$ and ${\bf U}_{l,n,{\bf k}}=
    \langle \ell,\mathbf k;0|{\bf U}|n,\mathbf k;0\rangle$. Thus ${\bf d}_{mn}^{(1)}$ is a vector coefficient describing how the dipole
matrix element changes under a polar distortion along different directions.
It can be obtained from first-principles calculations by evaluating the
change of optical matrix elements with respect to the components of
${\bf Q}$.

\subsection{Block diagonalization}

The SHG process involves a two-photon transition from occupied valence
states to empty conduction states through virtual intermediate states. The full Hamiltonian can therefore be written in block form,
\begin{equation}
H=
\begin{pmatrix}
H_{PP}&H_{PQ}\\
H_{QP}&H_{QQ}
\end{pmatrix}.
\end{equation}
Here,  $H_{PP}$ 
describes the direct valence-conduction sector, whereas $H_{QQ}$ contains all virtual intermediate states. To obtain an effective Hamiltonian within the $P$ subspace, we perform a
unitary transformation~\cite{PhysRev.149.491},
\begin{equation}
H_{\rm eff}
=
e^SHe^{-S},
\end{equation}
where $S^\dagger=-S$. Expanding the transformed Hamiltonian,
\begin{align}
H_{\rm eff}
=&
H+[S,H]
+\frac12[S,[S,H]]
+\cdots .
\end{align}
The generator $S$ is chosen to eliminate the coupling between the
$P$ and $Q$ subspaces,
\begin{equation}
H_{PQ}+[S,H_0]_{PQ}=0 .
\end{equation}
For the matrix elements connecting the two subspaces, 
$S_{mn}
=
\frac{V_{mn}}
{\varepsilon_m-\varepsilon_n}$, where $m\in P$ and $n\in Q$. Keeping terms up to second order in the optical coupling, the effective
Hamiltonian inside the $P$ subspace becomes~\cite{PhysRev.149.491}
\begin{equation}
H_{\rm eff}
=
H_{PP}
+
\frac12P[S,V]P .
\end{equation}
The effective coupling between a valence state $v$ and a conduction state
$c$ is therefore 
\begin{align}
\left(H_{\rm eff}^{(2)}\right)_{vc}
=&
\frac12
\sum_n
\left(
S_{vn}V_{nc}
-
V_{vn}S_{nc}
\right)\nonumber\\
=&
\frac12
\sum_n
V_{vn}V_{nc}
\left[
\frac1{\varepsilon_v-\varepsilon_n}
+
\frac1{\varepsilon_c-\varepsilon_n}
\right].
\end{align}
Substituting the optical matrix elements, 
$V_{mn}
=
-d_{mn}^{i}E_i$, we obtain
\begin{align}
\left(H_{\rm eff}^{(2)}\right)_{vc}
=
\sum_n
d_{vn}^{j}
d_{nc}^{k}
\Lambda_{vnc}
E_j(\omega)E_k(\omega),
\end{align}
where 
$\Lambda_{vnc}
=
\frac12[
\frac1{\varepsilon_v-\varepsilon_n}
+
\frac1{\varepsilon_c-\varepsilon_n}]$. This expression represents the complete two-photon optical vertex including
all virtual electronic bands.

\subsection{Second-order interband coherence}

After integrating out the virtual intermediate states, the remaining optical
response is described by the effective interaction between occupied valence
bands and empty conduction bands. The second-order interband density matrix obeys the equation of motion~\cite{Haug1994,Wu2010SpinDynamics}
\begin{align}
i\hbar
\frac{d}{dt}
\rho_{cv}^{(2)}(\mathbf{k},t)
=
\left(
\varepsilon_{c\mathbf{k}}
-
\varepsilon_{v\mathbf{k}}
-i\hbar\gamma
\right)
\rho_{cv}^{(2)}(\mathbf{k},t)\nonumber\\
+
f_{vc}(\mathbf{k})
d_{vc}^{i}(\mathbf{k},{\bf Q})
\mathcal F_{vc}^{jk}
(\mathbf{k},{\bf Q})
e^{-2i\omega t},
\end{align}
where 
$f_{vc}(\mathbf{k})
=
f_{v\mathbf{k}}
-
f_{c\mathbf{k}}$ is the occupation difference between the valence and conduction states. For an insulating system, 
$f_{vc}(\mathbf{k})\simeq 1$. Taking the stationary solution~\cite{Wu2010SpinDynamics,Haug1994}
\begin{equation}
\rho_{cv}^{(2)}(\mathbf{k},t)
=
\rho_{cv}^{(2)}
(\mathbf{k},2\omega)
e^{-2i\omega t},
\end{equation}
we obtain
\begin{align}
\rho_{cv}^{(2)}
(\mathbf{k},2\omega)
=
\frac{
f_{vc}(\mathbf{k})
d_{vc}^{i}(\mathbf{k},{\bf Q})
\mathcal F_{vc}^{jk}
(\mathbf{k},{\bf Q})
}
{
\varepsilon_{c\mathbf{k}}
-
\varepsilon_{v\mathbf{k}}
-
2\hbar\omega
-
i\hbar\gamma
}.
\end{align}
The second-harmonic polarization is generated by the induced interband
coherence~\cite{Haug1994,kittel1963quantum},
\begin{equation}
P_i^{(2)}
=
\sum_{\mathbf{k}}
\sum_{v,c}
d_{cv}^{i}(\mathbf{k},{\bf Q})
\rho_{vc}^{(2)}
(\mathbf{k},2\omega).
\end{equation}
\begin{widetext}
Substituting the above expression gives~\cite{Haug1994}
\begin{align}
P_i^{(2)}(2\omega)
=&
E_j(\omega)E_k(\omega)
\sum_{\mathbf{k}}
\sum_{v\in{\rm occ}}
\sum_{c\in{\rm unocc}}
\frac{
f_{vc}
d_{cv}^{i}
d_{vn}^{j}
d_{nc}^{k}
\Lambda_{vnc}
}
{
\varepsilon_{c}
-
\varepsilon_{v}
-
2\hbar\omega
-
i\hbar\gamma
}.
\end{align}
Therefore, 
$P_i^{(2)}(2\omega)
=
\chi_{ijk}^{(2)}({\bf Q})
E_j(\omega)E_k(\omega)$, 
with the electronic second-order susceptibility
\begin{align}
\chi_{ijk}^{(2)}({\bf Q})
=&
\sum_{\mathbf{k}}
\sum_{v\in{\rm occ}}
\sum_{c\in{\rm unocc}}
\sum_n
\frac{
f_{vc}(\mathbf{k})
d_{cv}^{i}(\mathbf{k},{\bf Q})
d_{vn}^{j}(\mathbf{k},{\bf Q})
d_{nc}^{k}(\mathbf{k},{\bf Q})
\Lambda_{vnc}(\mathbf{k})
}
{
\varepsilon_{c\mathbf{k}}
-
\varepsilon_{v\mathbf{k}}
-
2\hbar\omega
-
i\hbar\gamma
}.
\end{align}
\end{widetext}

\subsection{Effect of inversion-symmetry breaking distortion}

We next consider the dependence of the SHG susceptibility on the polar
distortion ${\bf Q}$. For the centrosymmetric reference structure, ${\bf Q}=0$, the second-order nonlinear susceptibility vanishes because the electric
dipole contribution is forbidden by inversion symmetry,
\begin{equation}
\chi_{ijk}^{(2)}(0)=0 .
\end{equation}
When inversion symmetry is broken by a finite polar distortion, the optical
matrix elements acquire a linear correction,

\begin{equation}
d_{mn}^{i}({\bf Q})
=
d_{mn}^{i,(0)}
+
Q_{\alpha}
d_{mn,\alpha}^{i,(1)}
+
O(Q^2).
\end{equation}
Consequently, the susceptibility can be expanded as
\begin{equation}
\chi_{ijk}^{(2)}({\bf Q})
=
\chi_{ijk}^{(2)}(0)
+
Q_\alpha
\chi_{ijk,\alpha}^{(2,1)}
+
O(Q^2).
\end{equation}
Because 
$\chi_{ijk}^{(2)}(0)=0 $, the leading contribution is
\begin{equation}
\chi_{ijk}^{(2)}({\bf Q})
=
Q_\alpha
\chi_{ijk,\alpha}^{(2,1)} .
\end{equation}
The coefficient describing the linear coupling between the polar distortion
and electronic SHG is
\begin{align}
&\chi_{ijk,\alpha}^{(2,1)}
=
\sum_{\mathbf{k}}
\sum_{v,c,n}
\frac{
f_{vc}
\Lambda_{vnc}
}
{
\varepsilon_c-\varepsilon_v
-
2\hbar\omega
-
i\hbar\gamma
}\Big[
d_{cv,\alpha}^{i,(1)}
d_{vn}^{j,(0)}
d_{nc}^{k,(0)}\nonumber\\
&
+
d_{cv}^{i,(0)}
d_{vn,\alpha}^{j,(1)}
d_{nc}^{k,(0)}
+
d_{cv}^{i,(0)}
d_{vn}^{j,(0)}
d_{nc,\alpha}^{k,(1)}
\Big].
\end{align}

Therefore, the electronic SHG polarization becomes
\begin{equation}
P_i^{(2)}(2\omega)
=
Q_\alpha
\chi_{ijk,\alpha}^{(2,1)}
E_jE_k .
\end{equation}
Finally, the SHG intensity is
\begin{equation}
I_{2\omega}
\propto
|P^{(2)}(2\omega)|^2 ,
\end{equation}
and therefore
\begin{equation}
I_{2\omega}
\propto
\left|
Q_\alpha
\chi_{ijk,\alpha}^{(2,1)}
E_jE_k
\right|^2 .
\end{equation}
Thus, within the complete all-band microscopic theory, the electronic SHG
susceptibility is linearly proportional to the inversion-symmetry-breaking
polar distortion, 
$\chi^{(2)}\propto {\bf Q}$, while the measured SHG intensity follows 
$I_{2\omega}\propto |{\bf Q}|^2$.

\section{Symmetry analysis}

We now analyze the symmetry content of the above result. The polar distortion
${\bf Q}$ is odd under inversion:
\begin{equation}
    {\cal P}:{\bf Q}\rightarrow -{\bf Q}.
\end{equation}
The electric field is also odd:
\begin{equation}
    {\cal P}:{\bf E}\rightarrow -{\bf E}.
\end{equation}
The second-order polarization satisfies
\begin{equation}
    P^{(2)}
    =
    \chi^{(2)}({\bf Q})E^2 .
\end{equation}
Under inversion,
\begin{equation}
    P^{(2)}\rightarrow -P^{(2)},
    \qquad
    E^2\rightarrow E^2 .
\end{equation}
Therefore the nonlinear susceptibility must obey
\begin{equation}
    \chi^{(2)}(-{\bf Q})
    =
    -\chi^{(2)}({\bf Q}).
\end{equation}
Hence $\chi^{(2)}$ is an odd function of the inversion-breaking vector order
parameter:
\begin{equation}
    \chi^{(2)}({\bf Q})
    =
    {\bf Q}\cdot\boldsymbol{\chi}^{(2)}_1
    +
    O(Q^3).
\end{equation}
This symmetry constraint explains why the centrosymmetric limit gives
\begin{equation}
    \chi^{(2)}({\bf Q}=0)=0,
\end{equation}
and why the leading electronic SHG response generated by inversion-breaking
orbital mixing is linear in ${\bf Q}$.

\section{Parity analysis of the dipole products}
\label{PS}

We now analyze the parity structure~\cite{sakurai2020modern} of the dipole products entering
$\chi^{(2)}_0$ and $\boldsymbol{\chi}^{(2)}_1$. In the centrosymmetric
reference structure, the electronic states may be assigned inversion
eigenvalues
\begin{equation}
    {\cal P}|n,\mathbf k;0\rangle
    =
    p_n |n,-\mathbf k;0\rangle,
    \qquad
    p_n=\pm 1 .
\end{equation}
The electric dipole operator is odd under inversion. Therefore the
zeroth-order dipole matrix element satisfies
\begin{equation}
    d_{mn}^{(0)}(-\mathbf k)
    =
    -p_m p_n d_{mn}^{(0)}(\mathbf k).
\end{equation}
Thus, a nonzero centrosymmetric dipole matrix element requires
\begin{equation}
    p_m p_n=-1 .
\end{equation}
Thus $d_{mn}^{(0)}$ connects states with opposite inversion parity.

The first-order correction ${\bf d}_{mn}^{(1)}$ contains one insertion of
the polar-distortion perturbation ${\bf U}$. Since ${\bf U}$ is odd under
inversion, ${\bf d}_{mn}^{(1)}$ contains the product of two odd operators:
the dipole operator and the polar mixing perturbation. Therefore it has even
parity overall and satisfies
\begin{equation}
    {\bf d}_{mn}^{(1)}(-\mathbf k)
    =
    p_m p_n
    {\bf d}_{mn}^{(1)}(\mathbf k).
\end{equation}
A nonzero ${\bf d}_{mn}^{(1)}$ therefore requires
\begin{equation}
    p_m p_n=+1 .
\end{equation}
Thus the inversion-breaking correction to the dipole matrix element connects
states with the same inversion parity.

We first consider the product appearing in $\chi^{(2)}_0$,
\begin{equation}
    T_0(\mathbf k)
    =
    d_{vc}^{(0)}(\mathbf k)
    d_{vn}^{(0)}(\mathbf k)
    d_{nc}^{(0)}(\mathbf k).
\end{equation}
Each factor is a zeroth-order dipole matrix element. Under
$\mathbf k\to -\mathbf k$, one obtains
\begin{align}
    T_0(-\mathbf k)
    &=
    \left[-p_vp_c\right]
    \left[-p_vp_n\right]
    \left[-p_np_c\right]
    T_0(\mathbf k)
   =
    -T_0(\mathbf k).
\end{align}
Therefore $T_0(\mathbf k)$ is odd under inversion. Since the energy
denominators and occupation factors are even functions of $\mathbf k$ in the
centrosymmetric reference structure, the full Brillouin-zone sum vanishes:
\begin{equation}
    \sum_{\mathbf k} T_0(\mathbf k)=0.
\end{equation}
Equivalently, from the selection-rule viewpoint, the simultaneous conditions
\begin{equation}
    p_vp_c=-1,
    \qquad
    p_vp_n=-1,
    \qquad
    p_np_c=-1
\end{equation}
cannot all be satisfied, because multiplying the last two equations gives
$p_vp_c=+1$, in contradiction with the first one. Hence the purely
centrosymmetric product cannot generate a nonzero bulk electric-dipole SHG
response.

We next consider the three terms linear in the polar distortion:
\begin{align}
    {\bf T}_1(\mathbf k)
    &=
    {\bf d}_{vc}^{(1)} d_{va}^{(0)} d_{ac}^{(0)}
    +
    d_{vc}^{(0)} {\bf d}_{va}^{(1)} d_{ac}^{(0)}
    +
    d_{vc}^{(0)} d_{va}^{(0)} {\bf d}_{ac}^{(1)} .
\end{align}
The contribution to the susceptibility is
\begin{equation}
    {\bf Q}\cdot{\bf T}_1(\mathbf k).
\end{equation}
Each term contains one first-order dipole correction and two zeroth-order
dipole matrix elements. For example,
\begin{align}
    &{\bf d}_{vc}^{(1)}(-\mathbf k)
    d_{va}^{(0)}(-\mathbf k)
    d_{ac}^{(0)}(-\mathbf k)
    =
    \left[p_vp_c\right]
    \left[-p_vp_a\right]
    \left[-p_ap_c\right]\nonumber\\
    &\times[
    {\bf d}_{vc}^{(1)}(\mathbf k)
    d_{va}^{(0)}(\mathbf k)
    d_{ac}^{(0)}(\mathbf k)]=
    {\bf d}_{vc}^{(1)}(\mathbf k)
    d_{va}^{(0)}(\mathbf k)
    d_{ac}^{(0)}(\mathbf k).
\end{align}
Thus this term is even under inversion. The same argument applies to the
other two terms,
\begin{equation}
    d_{vc}^{(0)} {\bf d}_{va}^{(1)} d_{ac}^{(0)},
    \qquad
    d_{vc}^{(0)} d_{va}^{(0)} {\bf d}_{ac}^{(1)} .
\end{equation}
Therefore ${\bf T}_1(\mathbf k)$ is even under $\mathbf k\to-\mathbf k$
and is not forced to vanish by inversion symmetry.

The same conclusion can be seen from the parity selection rules. For the first
term,
\begin{equation}
    {\bf d}_{vc}^{(1)} d_{va}^{(0)} d_{ac}^{(0)}\neq 0
\end{equation}
requires
\begin{equation}
    p_vp_c=+1,
    \qquad
    p_vp_a=-1,
    \qquad
    p_ap_c=-1,
\end{equation}
which are mutually consistent. Similarly, the second term requires
\begin{equation}
    p_vp_c=-1,
    \qquad
    p_vp_a=+1,
    \qquad
    p_ap_c=-1,
\end{equation}
and the third term requires
\begin{equation}
    p_vp_c=-1,
    \qquad
    p_vp_a=-1,
    \qquad
    p_ap_c=+1.
\end{equation}
All three sets of conditions are allowed. Hence the linear-in-${\bf Q}$
correction opens parity-allowed optical pathways that are forbidden in the
centrosymmetric limit.

Therefore the centrosymmetric contribution satisfies
\begin{equation}
    \chi^{(2)}({\bf Q}=0)=0,
\end{equation}
whereas the leading inversion-breaking contribution is
\begin{equation}
    \chi^{(2)}({\bf Q})
    =
    {\bf Q}\cdot\boldsymbol{\chi}^{(2)}_1 .
\end{equation}
This explicitly shows that the electronic SHG response is activated by
inversion-symmetry-breaking orbital mixing and is linear in the polar
distortion vector ${\bf Q}$ at the susceptibility level.

\section{Effective theory for THz-pump optical-SHG-probe}

Pump-probe spectroscopy provides a powerful approach to investigate and
control the nonequilibrium evolution of polar lattice distortions by using an ultrafast pump pulse to drive a
system away from equilibrium and a delayed probe pulse to monitor its
subsequent evolution~\cite{li2023terahertz,cheng2023terahertz,74d5-4hsw,Basini2024,Nova2019,Fechner2024,RevModPhys.93.041002}. In displacive ferroelectrics, THz-pump pulses are
often employed because their frequency is close to the characteristic energy
scale of polar lattice dynamics. The soft polar phonon modes typically lie
in the THz frequency range~\cite{rowley2014ferroelectric,Yang2026,cowley1980structural,cochran1981soft,cochran1961crystal,cowley1996phase,cowley1965theory,cochran1969dynamical,cochran1960crystal,Yang2025,Yang2026PDW}, allowing the THz optical field to resonantly
excite coherent lattice distortions. The driven polar mode can therefore be
described by a time-dependent inversion-symmetry-breaking lattice
displacement ${\bf Q}(t)$. To probe the transient response induced by the THz excitation while avoiding
direct interference with the lattice dynamics, the subsequent measurement is
often performed using an optical SHG probe~\cite{li2023terahertz,cheng2023terahertz,74d5-4hsw,Basini2024,Nova2019,Fechner2024,RevModPhys.93.041002}, typically with a wavelength of
$800~\mathrm{nm}$, corresponding to a photon energy of
$\hbar\omega\approx1.55~\mathrm{eV}$ and a two-photon energy of
$2\hbar\omega\approx3.1~\mathrm{eV}$.  This optical frequency is much higher
than the characteristic frequency of soft phonon modes, such that the probe
field cannot directly follow or drive the ionic motion. Thus, the optical
SHG response provides a sensitive probe of the electronic structure modified
by the THz-driven polar distortion.

Although the optical excitation directly couples to the electronic system,
displacive ferroelectrics are typically insulating materials with well-defined
valence and conduction bands. Therefore, the optical SHG response is dominated
by virtual interband electronic transitions rather than free-carrier
dynamics~\cite{Haug1994,kittel1963quantum}. The polar distortion controls the SHG response indirectly through
the reconstruction of electronic Bloch wave functions, which modifies the
optical dipole matrix elements and activates inversion-symmetry-breaking
optical pathways.

As derived above, the electronic SHG susceptibility can be expressed as
\begin{align}
\chi_{ijk}^{(2)}({\bf Q})
=
\sum_{{\bf k},v,c,n}
\frac{
f_{vc}\,
d_{vc}^{i}
d_{vn}^{j}
d_{nc}^{k}
\Lambda_{vnc}
}{
\varepsilon_{c{\bf k}}
-
\varepsilon_{v{\bf k}}
-
2\hbar\omega
-
i\hbar\gamma
},
\end{align}
where $v$, $c$, and $n$ denote occupied, empty, and intermediate Bloch
bands, respectively. Here, $f_{vc}=f_v-f_c$ is the occupation difference (or Pauli blocking factor) between the initial
valence band and final conduction band states~\cite{Haug1994,kittel1963quantum}. Expanding around the centrosymmetric structure and applying inversion
symmetry gives the leading contribution
\begin{equation}
\chi_{ijk}^{(2)}({\bf Q})
=
Q_\alpha
\chi_{ijk,\alpha}^{(2,1)}
+
O(Q^3),
\end{equation}
where $\chi_{ijk,\alpha}^{(2,1)}$ is the linear SHG susceptibility response
to a polar distortion along the $\alpha$ direction. Importantly, these
effective coupling coefficients are fully determined by the distortion-induced
changes of Bloch wave functions and optical matrix elements, and can be
directly evaluated from first-principles calculations by computing the
electronic structure and optical response of slightly distorted crystal
structures.

On the other hand, the THz electric field couples directly to the polar lattice mode and
drives a time-dependent inversion-breaking distortion. The dynamics of the
polar coordinate is described by~\cite{cheng2023terahertz,74d5-4hsw}
\begin{equation}\label{ddd}
\ddot{Q}_{\alpha}(t)
+
\gamma_Q\dot{Q}_{\alpha}(t)
+
\Omega_{\alpha}^{2}Q_{\alpha}(t)
=
\frac{z_\alpha^*}{m_\alpha}
E_{{\rm THz},\alpha}(t),
\end{equation}
where $\Omega_\alpha$, $\gamma_Q$, $z_\alpha^*$, and $m_\alpha$ denote the
polar-mode frequency, damping coefficient, mode effective charge, and
effective mass, respectively. The THz pulse therefore produces a transient polar distortion
$Q_\alpha(t)$, which dynamically modulates the electronic SHG susceptibility,
\begin{equation}
\chi_{ijk}^{(2)}(t)
=
Q_\alpha(t)
\chi_{ijk,\alpha}^{(2,1)} .
\end{equation}

Consequently, the time-dependent SHG signal becomes
\begin{equation}\label{SHGSHG}
I_{2\omega}(t)
\propto
\left|
Q_\alpha(t)
\chi_{ijk,\alpha}^{(2,1)}
E_j(\omega)E_k(\omega)
\right|^2 .
\end{equation}
Therefore, the THz field provides an external control knob for the nonlinear
optical response by coherently manipulating the polar lattice dynamics. This
effective theory establishes a direct connection between nonequilibrium
phonon motion and ultrafast electronic SHG, while the key material-dependent
parameters can be obtained from first-principles calculations.

\begin{figure}[h]
    \centering
    \includegraphics[width=\linewidth]{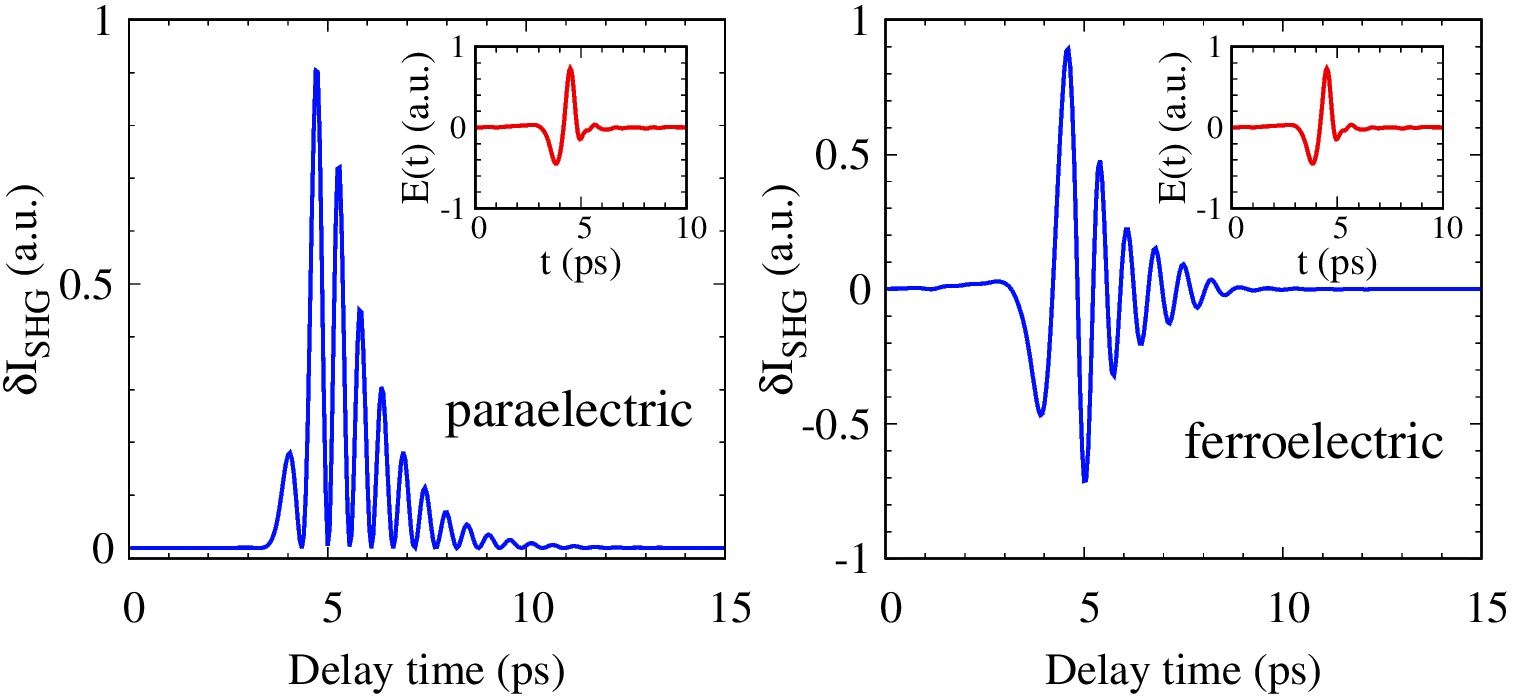}
    \caption{
    Simulated THz-pump optical-SHG-probe response in the paraelectric
    (left) and ferroelectric (right) phases obtained from Eqs.~(\ref{ddd}) and (\ref{SHGSHG}).
    The driving electric field is a single-cycle THz pulse, as shown in the
    inset, and the polar-mode frequency $\Omega_{\alpha}$ is chosen to be
    resonant with the THz pulse. In the paraelectric phase, $\delta I_{\rm SHG}$ remains positive and exhibits a frequency-doubled oscillation, whereas in the ferroelectric phase,
    $\delta I_{\rm SHG}$ oscillates around zero predominantly at the
    fundamental polar-mode frequency.
    }
    \label{fig:energy}
\end{figure}

To illustrate the distinct pump-probe responses in the paraelectric and
ferroelectric phases, we numerically solve the driven equation of motion for
the polar mode under a THz pulse, with the results shown in Fig.~\ref{fig:energy}.
The left panel in Fig.~\ref{fig:energy} shows the response in the paraelectric phase. Since the
equilibrium polar distortion vanishes, the THz pulse drives the polar
coordinate around $Q=0$. As a consequence of
$I_{\rm SHG}(t)\propto Q^2(t)$, the corresponding change in the SHG intensity,
$\Delta I_{\rm SHG}(t)$, remains positive and exhibits oscillations at twice
the frequency of the driven polar mode. In addition to this frequency-doubled
oscillation, a finite positive background is present because the two opposite
polar distortions, $+Q$ and $-Q$, give the same SHG intensity. The right panel in Fig.~\ref{fig:energy} shows the corresponding response in the ferroelectric phase.
In this case, the polar coordinate oscillates around a finite equilibrium
distortion $Q_0$. The THz-driven displacement therefore periodically increases
and decreases the magnitude of the ferroelectric distortion relative to its
equilibrium value. Accordingly, $\Delta I_{\rm SHG}(t)$ exhibits an oscillatory
response with alternating enhancement and suppression of the SHG intensity.
Its dominant oscillation follows the frequency of the driven polar mode,
rather than the frequency-doubled behavior characteristic of the paraelectric
phase. These characteristic behaviors of paraelectric and ferroelectric phase are consistent with experimental observations
of THz-pump optical-SHG-probe measurements in paraelectric and ferroelectric
systems~\cite{74d5-4hsw}.

\section{Discussion}

The microscopic mechanism revealed in this work provides a different
perspective on ferroelectric SHG beyond the conventional
symmetry-based description. Symmetry analysis determines whether a nonzero
second-order susceptibility is allowed, but it does not specify how the
electronic nonlinear optical response emerges from a structural distortion.
Here we show that the essential microscopic process is the reconstruction of
electronic Bloch states induced by the inversion-breaking polar displacement.
The lattice degree of freedom does not directly participate in the optical
transition; it modifies the electronic wave functions and optical
transition matrix elements, thereby controlling the nonlinear optical vertex.

An important consequence of this picture is that the ferroelectric order
parameter acts as a microscopic control parameter for electronic nonlinear
optics. In the centrosymmetric phase, the electric-dipole SHG response is
forbidden because the optical pathways cancel under inversion symmetry. The
polar distortion removes this cancellation by mixing electronic states with
different inversion characters and activating parity-allowed virtual
transitions. Therefore, the emergence of SHG is not merely a consequence of
symmetry breaking, but results from distortion-induced modifications of the
electronic structure. This provides a microscopic interpretation of the
frequently observed correlation between ferroelectric polarization and SHG
intensity.

The framework also clarifies the role of THz-driven phonon dynamics in
ultrafast nonlinear optical control. In conventional pump-probe descriptions,
the THz field is viewed as a means to manipulate the structural order
parameter, while the connection to the optical response is often introduced
phenomenologically. The theory reveals a direct microscopic
link:
\[
E_{\rm THz}(t)\rightarrow Q(t)\rightarrow
|\psi_n({\bf Q})\rangle\rightarrow
\chi^{(2)}(t).
\]
The coherent oscillation of a polar phonon therefore provides a dynamical
modulation of the electronic SHG susceptibility. This mechanism suggests that
phonon engineering can be used as a general strategy to control nonlinear
optical responses on ultrafast time scales. The formulation also provides a practical route toward material-specific
predictions. The coupling tensor $\chi_{ijk,\alpha}^{(2,1)}
={\partial \chi_{ijk}^{(2)}}/{\partial Q_\alpha}$ can be evaluated from first-principles calculations~\cite{PhysRevB.53.10751,PhysRevB.57.3905,PhysRevB.71.125107} by computing the evolution
of electronic wave functions and optical matrix elements under small polar
displacements. This enables quantitative comparison with THz-pump
optical-SHG-probe experiments and provides a microscopic design principle for
materials with strongly tunable nonlinear optical responses.

Several previous phenomenological approaches~\cite{Ross2026,Ross2026KNbO3,PhysRevB.111.085109,Li2026THzNanotwin} have attempted to describe ferroelectric optical responses by introducing an effective electronic polarization field within a Landau framework. In these treatments, the electronic polarization is modeled in analogy with a lattice or phonon polarization coordinate and is assigned a bosonic-like dynamical equation of motion, typically of the Klein-Gordon form~\cite{peskin2018introduction},
\begin{equation}
\mu_e \frac{\partial^2 P_e}{\partial t^2}
+
\gamma_e \frac{\partial P_e}{\partial t}
+
\frac{\partial F}{\partial P_e}
=
E(t).
\end{equation}
Such a description effectively treats the electronic polarization as a collective dynamical coordinate analogous to an optically active phonon mode. However, this assumption conflates
the fundamentally different natures of lattice and electronic degrees of
freedom. While the lattice polarization corresponds to a collective ionic
displacement and can be described by bosonic order-parameter dynamics, the
polarization induced by electronic excitations originates from fermionic
interband transitions and represents a quantum mechanical response function~\cite{Haug1994,kittel1963quantum,PhysRevB.53.10751,PhysRevB.57.3905,PhysRevB.71.125107}
rather than an independent dynamical field. In particular, the above phenomenological equation leads to a Lorentz-oscillator-type
response~\cite{pekker2015amplitude,Yang2024,matsunaga2013higgs,Fang2026,matsunaga2014light,Yang2023,gpbp-qhp9,shimano2020higgs,PhysRevB.46.15085,Torchinsky2013},
\begin{equation}
    \chi_{\rm KG}(\omega)
    =
    \frac{1}
    {\Omega_e^2-\omega^2-i\gamma_e\omega},
\end{equation}
which describes a bosonic collective-mode resonance. In contrast, the
microscopic electronic polarization, determined by the quantum mechanical
interband response of electrons, is well established in the literature~\cite{Haug1994,kittel1963quantum,PhysRevB.52.14636,PhysRevB.48.11705},
\begin{equation}
    P_i(\omega)
    =
    \epsilon_0\chi_{ij}^{(1)}(\omega)E_j(\omega),
\end{equation}
where
\begin{equation}
    \chi_{ij}^{(1)}(\omega)
    =
    \frac{1}{\epsilon_0}
    \sum_{m,n,\mathbf{k}}
    \frac{
    (f_{m\mathbf{k}}-f_{n\mathbf{k}})  d^i_{mn}(\mathbf{k})
    d^j_{nm}(\mathbf{k})
    }
    {\varepsilon_{n\mathbf{k}}
    -
    \varepsilon_{m\mathbf{k}}
    -
    \hbar\omega-i\eta}.
\end{equation}
Therefore, the optical response should be derived from the evolution of
electronic wave functions and optical transition matrix elements, rather than
from an effective equation of motion for an electronic polarization field. Such a bosonic-field description therefore does not, in general, capture the actual microscopic physics of electronic excitations, particularly their interband nature, and may consequently lead to a qualitatively different frequency dependence of the optical response.

\bibliography{ref.bib}

\end{document}